\documentclass[prb,aps,twocolumn,superscriptaddress]{revtex4}
\usepackage{bm}
\usepackage{graphicx}
\usepackage{amsfonts}
\usepackage{amsmath}
\usepackage{amssymb}
\newcommand {\be}{\begin{equation}}
\newcommand {\ee}{\end{equation}}
\newcommand {\bea}{\begin{eqnarray}}
\newcommand {\eea}{\end{eqnarray}}

\begin{document}

\date{\today}
\title{ Superfluidity in asymmetric electron-hole systems}
\author{ Ilya Grigorenko}
\affiliation{ Physics Department, New York City College of Technology, The City University
of New York, Brooklyn, NY 11201, USA}
\author{Roman Ya. Kezerashvili }
\affiliation{ Physics Department, New York City College of Technology, The City University
of New York, Brooklyn, NY 11201, USA}
\affiliation{The Graduate School and University Center, The City University of New York, New York, New York 10016, USA}

\begin{abstract}
The pairing in a system of electrons and holes in two spatially separated
parallel planes is studied in the case of electron-hole asymmetry caused by
the difference in the carriers masses and their chemical potentials.
It is predicted that the system may exhibit two critical temperatures for some range of the asymmetry parameters.
The lower critical temperature corresponds to the superfluid transition induced by thermal fluctuations. 
It is found that the superfluid state is possible in a wide range of the asymmetry parameters,
because the asymmetries can effectively compensate each other.  In the asymmetric system a coexistence of the
normal and superfluid states is possible even at zero temperature. 
\end{abstract}

\maketitle






Superfluid transition in  systems of two parallel planes with spatially
separated electrons and holes has been a subject of interest for many years 
\cite{Lozovik,VM,BJL,JBL,KGB}. Drag currents and zero charge dipolar current
were predicted in such systems \cite{VM,BJL}. In spatially separated system
of electrons and holes in two parallel planes the chemical potentials and
the mass of the positive and negative carriers can be controlled
independently, making them a good testbed for studies of the electron-hole
asymmetry on the formation of the superfluid state. It is common that in
chemically doped semiconductor systems the positive and negative charge
carriers may have very different effective masses. The ratio between the
electron and hole masses $m_{e}/m_{h}$ may vary, dependent on the material
and the doping levels \cite{eff_mass0}. Recently there were suggestions to
engineer the effective mass value \cite{eff_mass}. It was assumed that for a
given electron-hole mass ratio the superfluidity exists and the change of
this parameter would change the results only quantitatively \cite{VM}.
Besides, it was usually assumed that the most favorable condition for
pairing occurs at the equal concentrations of electrons and holes that for equal carrier masses
corresponds to the equal chemical potentials. Even a small carriers concentration
difference would lead to a destruction of the superfluid state, because for
nonequal carrier concentrations not all particles can be paired. It is
expected that the mismatch between the chemical potentials $\mu _{e}\neq \mu
_{h}$ may significantly reduce the critical temperature of the
superconducting transition. For sufficiently large mismatch between the
chemical potentials, comparable to the order parameter, the superfluid transition
may be impossible. This situation was extensively studied in the context of
asymmetric nuclear matter \cite{nuclear,nuclear0}, where the different number of
protons and neutrons results in different Fermi energies, and in cold Fermi
gases with two or more atomic species \cite{cold_gases}. In nuclear matter
the asymmetry comes from the different concentrations of protons and
neutron, while their masses are usually assumed identical \cite{nuclear1}.
The asymmetric nuclear matter may exhibit coexistence of paired and unpaired
components at zero temperature and two critical temperatures, that results
in superfluidity in some temperature range \cite{nuclear2,nuclear3}.
These studies predicted that even a tiny proton-neutron
asymmetry destroys the superfluid state, when the chemical potential
difference becomes comparable to the magnitude of the order parameter in the
symmetric system.

In this work we study an asymmetric electron-hole system, where the electrons reside in
a two-dimensional layer and the holes in another two-dimensional layer,
which can exhibit superfluidity in a very wide range of the electron-hole
mass and chemical potential asymmetry. We found a range of
the parameters, where the system has two critical temperatures, and
superfluidity exists in a finite temperature interval. We show that the superfluid state
survives because the electron-hole mass asymmetry can be effectively
compensated by the chemical potential difference. We also predict that at
zero temperature with the change of the asymmetry parameters the system may
undergo the quantum phase transition of the first order. 

Let us consider a system of two parallel planes separated by a dielectric
with spatially separated electrons and holes confined on each plane. 
On one hand even in the case of perfectly equal carrier concentrations and zero temperature, the Fermi energy levels may be
significantly different due to the difference in the carriers masses. And on
the other hand, for unequal masses and equal Fermi energies the difference
in concentrations of the carriers in the planes will result in a
considerable amount of the carriers, which can not find their pairs. We are
presenting the study how these two types of asymmetry interfere with each other
and affect the formation of the superfluid state. Let us assume the carrier
concentrations in the system are $\rho _{e}$ and $\rho _{h}$ for electrons
and holes, respectively. The effective Hamiltonian can be written as \cite%
{Lozovik} 
\begin{eqnarray}
H_{eff} &=&\sum_{\mathbf{p}}(\xi_{\mathbf{p}}-\mu_{e})b_{\mathbf{p}%
}^{\dagger }b_{\mathbf{p}}+\sum_{\mathbf{{p^{\prime }}}}(\xi_{\mathbf{{%
p^{\prime }}}}^{\prime }-\mu_{h})a_{\mathbf{p^{\prime }}}^{\dagger }a_{%
\mathbf{p^{\prime }}}+  \notag  \label{Heff} \\
&&+\sum_{\mathbf{p}}\left[ \Delta _{\mathbf{p}}b_{\mathbf{p}}^{\dagger }a_{-%
\mathbf{p^{\prime }}}^{\dagger }+H.c.\right] ,
\end{eqnarray}%
where $a_{\mathbf{p^{\prime }}}$ is the operator of annihilation of a hole
on one plane, and $b_{\mathbf{p}}$ is the operator of annihilation of an
electron on another plane. The single-particle energies of electrons are
given by $\xi _{\mathbf{p}}=\frac{|\mathbf{p}|^{2}}{2m_{e}}=\frac{\hbar ^{2}|%
\mathbf{k}|^{2}}{2m_{e}}$. Similarly the single-particle energies of holes
are presented as $\xi _{\mathbf{p}}^{\prime }=\frac{{|\mathbf{p}}|^{2}}{%
2m_{h}}=\frac{\hbar ^{2}{|\mathbf{k}}|^{2}}{2m_{h}}$. The
masses of the electrons and holes $m_{e}$ and $m_{h}$ and
the chemical potentials $\mu_{e}$ and $\mu_{h}$ we treat as independent
parameters. In Eq.~(\ref{Heff}) the nonzero order parameter $\Delta _{%
\mathbf{p}}$ shows that the system is in the superfluid phase.
One can diagonalize Eq.~(\ref{Heff}) using Bogoliubov unitary transformations 
$a_{\bf{p}}=u_{\bf{p}} \alpha^\dagger_{-p}+v_{\bf{p}}
\beta^\dagger_{-p}$ and $b_{\bf{p}}=u_{\bf{p}} \beta_{\bf{p}}-v_{
\bf{p}} \alpha_{\bf{p}}$,  with the amplitudes $u_{\bf{p}}$ and $
v_{\bf{p}}$. The self-consistency condition for the order parameter has the form \cite
{Lozovik} 
\begin{eqnarray}  
\label{order_param}
\Delta_{\bf{p}}= \sum_{\bf{{p^{\prime }}}} V({\bf{p}}-{\bf{{
p^{\prime }}}}) u_{\bf{p^{\prime }}} v_{\bf{p^{\prime }}}(1-f(E_{+})-f(E_{-})), 
\end{eqnarray}
where $V(\bf{p})$ is the screened Coulomb attractive interaction between electrons and
holes. Here we use the notation $E_{\pm}=E\pm\eta_{\mathbf{p}}$, 
$E=\sqrt{\epsilon_{\mathbf{p}}^2+\Delta^2}$, 
$\epsilon_{\mathbf{p}}=(\xi_{\mathbf{p}}+\xi^{\prime }_{\mathbf{p}}-\mu_e-\mu_h)/2$, 
$\eta_{\mathbf{p}}=(\xi_{\mathbf{p}}-\mu_e-\xi^{\prime }_{\mathbf{p}}+\mu_h)/2$. 
In Eq.~(\ref{order_param}) the Fermi-Dirac distribution function is given
by $f(\epsilon) =[\exp(\epsilon/(k_{B}T)) + 1]^{-1}$, where $k_{B}$ is the
Boltzmann constant, and $T$ is temperature. The amplitudes $u_{\mathbf{p}}$, 
$v_{\mathbf{p}}$ are given by 
$u_{\mathbf{p}}^2=\frac{1}{2}(1+\frac{\epsilon_{\mathbf{p}}}{E}%
), v_{\mathbf{p}}^2=\frac{1}{2}(1-\frac{\epsilon_{\mathbf{p}}}{E%
}). $ 
It is convenient to introduce new variables. First, we introduce the mass
asymmetry parameter $\alpha=(m_e-m_h)/m_h$ and the chemical potential
difference $\delta \mu=\frac{\mu_e-\mu_h}{2}$. We also introduce the average
kinetic energy $\xi=(\xi_{\mathbf{p}}+\xi^{\prime }_{\mathbf{p}})/2$, the
average chemical potential $\mu^*=\frac{\mu_e+\mu_h}{2}$, and the effective
asymmetry variable $r=\alpha \mu^*+\delta \mu$. In the new variables the
quasiparticle energies can be represented as $E_{\pm}=\sqrt{%
(\xi-\mu^*)^2+\Delta^2}\pm(\alpha\xi+\delta\mu)$.

One can
use the result for the interaction potential obtained for the electron-hole
pairing in two parallel plane layer, given by Eq.~(7) in Ref.~\citenum{Lozovik}.
Assuming the constant attraction of the paired particles one can replace the
interaction term $V(\mathbf{p}-\mathbf{{p^{\prime }}})$ by a constant
effective interaction $U\propto \frac{e^{2}}{2\varepsilon \varepsilon
_{0}k_{F}}\exp (-D k_{F})$, where $k_{F}$ is the Fermi wavevector, $%
\varepsilon $ is the characteristic dielectric constant of the dielectric
between the planes, and  $D$ is the separation distance between the planes. 
The cut-off for the effective interaction is set by the
characteristic plasma frequency \cite{Lozovik}. To simplify the calculations
we assume that the order parameter is independent on the momentum: $\Delta _{%
\mathbf{p}}\equiv \Delta $. The chemical potentials for electrons $\mu _{e}$
and holes $\mu _{h}$ are coupled self-consistently to the given surface
densities $\rho _{e}$ and $\rho _{h}$ of the carriers : 
\begin{eqnarray}
\rho _{e} &=&\frac{2}{S}\sum_{\mathbf{p}}\left[ f(E_{+})u_{\mathbf{p}
}^{2}+(1-f(E_{-}))v_{\mathbf{p}}^{2}\right] ,  \notag  \label{dens} \\
\rho _{h} &=&\frac{2}{S}\sum_{\mathbf{p}}\left[ f(E_{-})u_{\mathbf{p}
}^{2}+(1-f(E_{+}))v_{\mathbf{p}}^{2}\right] ,
\end{eqnarray}
where $S$ is the unit of the surface area. The chemical potentials $\mu _{e}$
and $\mu _{h}$, and the gap $\Delta $ can be determined by the simultaneous
solution of Eqs.~(\ref{order_param}) and (\ref{dens}). 
If the chemical potentials are much larger than the order parameter, one can
neglect their changes due to the superfluid transition. This significantly
simplifies the calculations. 
Without losing generality let us assume the density of electrons
is higher than the density of holes: $\rho _{e}>\rho _{h}$. The excesses
electrons can not find their pairs, and they may form the normal phase. This phase
occupies certain energy levels. Analyzing the density difference 
\begin{eqnarray}
\delta \rho =\rho _{e}-\rho _{h}=\frac{2}{S}\sum_{\mathbf{p}}\left[
f(E_{+})-f(E_{-}))\right] ,
\end{eqnarray}
one notes that in the limit of low temperatures $T\rightarrow 0$ the Fermi
distribution function becomes the step function $f(x)\rightarrow \theta (-x)$. Because of the algebraic structure only one out of the quasiparticle
energies $E_{+}$ and $E_{-}$ can take both positive and negative values, the
other quasiparticle energy keeps its sign. Let us assume $E_{+}>0$ for all values of the
momentum $\mathbf{p}$, and $\theta (-E_{+})\equiv 0$. In this case the
unpaired electron fraction occupies the energy levels within $E_{-}<0$.
Considering the inequality $E_{-}=\sqrt{\epsilon _{\mathbf{p}}^{2}+\Delta
^{2}}-\eta _{\mathbf{p}}
<0$, one can observe that the unpaired particles occupy the energy interval $%
[\xi _{1},\xi _{2}]$  
\begin{eqnarray}
\label{interval}
\xi_{1,2}=\mu^*+(\alpha r\pm \sqrt{r^{2}-g^{2}\Delta^{2}})g^{-2},
\end{eqnarray}%
where we use the notation $g=\sqrt{1-\alpha ^{2}}.$ In general case the interval 
$[\xi _{1},\xi _{2}]$ is asymmetric with respect to $\mu ^{\ast }$. If the  condition $\Delta \geq 
\frac{|r|}{g}$ is satisfied, then the superfluid
and unpaired normal phase may coexist even at zero temperature. The presence of unpaired component
partially blocks the pairing near the Fermi surface. The energy levels
occupied by the unpaired electrons are automatically excluded in Eq.~(\ref%
{order_param}). At low temperatures $T\rightarrow 0$ it becomes 
$1=V\sum_{\mathbf{p}}\frac{1-\theta (-E_{-})}{2E}$, 
i.e. only states with $E_{-}>0$ participate in the pairing.

In this interesting case the standard Bardeen-Cooper-Schrieffer variational ground state
wavefunction is not useful. To describe the asymmetric system one may use
the following variational wavefunction \cite{nuclear} 
\begin{eqnarray}
\label{psi}
\Psi
=\prod_{k_{i}}a_{k_{i}}^{+}%
\prod_{k_{j}}(u_{k_{j}}+u_{k_{j}}a_{k_{j}}^{+}a_{-k_{j}}^{+})|0>.
\end{eqnarray}%
The wavefunction (\ref{psi}) describes the paired electron-hole component, as well as
the unpaired component that in our case are the excess electrons. The
summation over $k_{i}$ and $k_{j}$ correspond to $E_{-}<0$ and $E_{-}>0$,
respectively. It is necessary to note that the condition $E_{-}<0$ may have no solutions
in the case of relatively small asymmetries. It means that while the
densities of holes and electrons are not exactly equal, they all still form
the superfluid condensate. 

In the case of infinite planes one replaces the summation over $%
\mathbf{p}$ by two dimensional integration. One can make a
standard change of variables in the self-consistency equation transforming from the
momentum  to the energy integration:
\begin{eqnarray}
\label{order_param_at_01}
1=N(0)V\int_{\mu^*-\xi_{c}}^{\mu^*+\xi_{c}} d\xi \frac{\theta(\xi_1-\xi)+\theta(\xi_2-\xi)}{2 E},
\end{eqnarray}%
where $\xi _{c}$ is the high-energy cut-off, $N(0)$ is the density of states
at the Fermi level, which in the case of two dimensional systems is a
constant $N(0)=m^*/\pi \hbar^{2},$ where $m^*$ is the reduced
mass the  electron-hole pair.
The expression on the left in Eq.~(\ref{order_param_at_01}) can be represented as $N(0)V\ln(\Delta/\Delta_0)$.
The self-consistency Eq.~(\ref{order_param_at_01}) has two possible solutions. 
The first solution is a constant gap
$\Delta=\Delta_0$, where $\Delta _{0}$ is the gap in the symmetric case $\alpha =\delta
\mu =0$  at zero temperature. The second non-trivial solution can be found
analytically in some limiting cases. If one assumes $\mu ^{\ast }\gg \xi
_{c}\gg \Delta $, then $-\xi _{c}+\sqrt{(\xi _{c})^{2}+\Delta ^{2})}\approx 
\frac{1}{2}\frac{\Delta ^{2}}{\xi _{c}}$ and $(\xi _{c}+\sqrt{(\xi
_{c})^{2}+\Delta ^{2})}\approx 2\xi _{c}$. With these simplifications the
self-consistency condition becomes 
\begin{eqnarray}
\label{eq8}
1=N(0)V\ln \left[ \frac{4\xi_{c}^{2}(\xi _{2}-\mu^*-\sqrt{(\xi
_{2}-\mu^*)^{2}+\Delta ^{2}})}{\Delta^{2}(\xi_{1}-\mu^*-%
\sqrt{(\xi_{1}-\mu^*)^{2}+\Delta^{2}})}\right] .
\end{eqnarray}%
Using that $\sqrt{(\xi_{i}-\mu ^{\ast })^{2}+\Delta ^{2}}=\alpha \xi_{i}+\delta 
\mu $, $i=1,2$ one can further simplify Eq.~(\ref{eq8}):
\begin{eqnarray}
\label{simple_delta}
(\xi _{2}-\mu ^{\ast }+\alpha \xi _{2}+\delta \mu )(\xi _{1}-\mu ^{\ast
}-\alpha \xi _{1}-\delta \mu )=-1/\Delta_0^2.
\end{eqnarray}%
Substituting the expressions for $\xi_{1}$ and $\xi_{2}$ from Eq.~(\ref{interval}) 
into Eq (\ref{simple_delta}) one can find an analytical
expression for the order parameter 
\begin{eqnarray}
\label{delta_sol}
\Delta =\Delta_{0}\sqrt{1-\frac{2|r|}{g \Delta_{0}}}.
\end{eqnarray}%
Now let us study the thermodynamical stability of the obtained solutions.
Following Ref.~14, we are using the theorem that for small variations of an external parameter
of the system the grand potential varies in the same way as the Hamiltonian: $\frac{\partial \Omega}{\partial V}=<\frac{\partial H}{\partial V}>$.
Using the expression for the Hamiltonian of
the system \cite{significant} it is possible to write the expression for the grand potential: 
\begin{eqnarray}
\label{omega}
\Omega =-\int \frac{dV}{V^2}\int d{\bf r} |\Delta({\bf r})|^2.
\end{eqnarray}%
One can make change of variables, using the result for the gap $\Delta_0$ in the symmetric case:
$\frac{d \Delta_0}{\Delta_0}=\frac{2 d V}{N(0) V^2}$.
For a homogeneous system the difference between the grand potential  of the superconducting and
the normal states is 
\begin{eqnarray}
\label{omega1}
\Omega-\Omega_0 =-\frac{N(0)}{2}\int_{\Delta_1}^{\Delta_0} |\Delta|^2 \frac{d\Delta_0}{\Delta_0},
\end{eqnarray}%
where $\Delta_1$  is the value of $\Delta_0$ corresponding to $\Delta=0$.
It is easy to compare the value of the grand potential 
for two solutions. By substituting the solution with the constant gap $\Delta=\Delta_0$ in Eq.~(\ref{omega1})
and integrating from $\Delta_1=0$ to $\Delta_0$ we obtain $-N(0)\Delta_0^2/4$. If we now substitute
Eq.~(\ref{delta_sol}) into Eq.~(\ref{omega1}), integrate from $\Delta_1=2|r|/g$ to $\Delta_0$ and compare
the obtained value with the previous one, we can conclude that the solution with the constant gap $\Delta=\Delta_0$
always gives the lower grand potential.

\begin{figure}
\includegraphics[width=6.cm,angle=0]{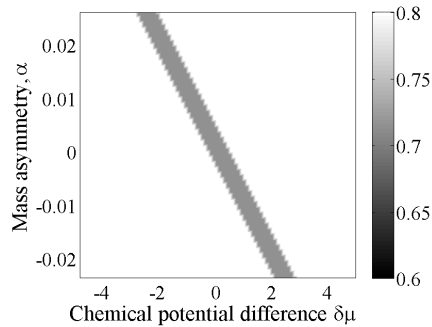} 
\caption{{\protect\footnotesize {The gap $\Delta$ as a function of the
mass asymmetry parameter $\protect\alpha$ and the chemical potential
difference $\protect\delta \protect\mu$ at zero temperature $T=0$.  Note that the line determined by the condition $r\equiv
\protect\alpha\protect\mu^*+\delta \protect\mu=0$ goes as the middle line of the ''stripe''. 
System parameters are $\protect\xi_c/\protect\mu^*=0.1$%
, $V/\protect\mu^*=1.5\times 10^{-3}$. }}}
\label{label_figure0}
\end{figure}

In Fig.~\ref{label_figure0} we plot the gap $\Delta$ as a function of the asymmetry parameters
$\alpha$ and $\Delta \mu$. For the asymmetry parameters  satisfying  the condition
$r\equiv\protect\alpha\protect\mu^*+\delta \protect\mu=0$ or in the close vicinity, the superfluid
state is possible and is characterized by an independent on the asymmetry parameters gap $\Delta=\Delta_0$.
For relatively big asymmetries the normal state with  $\Delta=0$ becomes thermodynamically favored,
because it results in lower values of the grand potential. In 
Fig.~\ref{label_figure0} it corresponds to the area outside of the stripe. 
The transition to the normal state is of the first order, since the
order parameter does not depend on the asymmetry parameters.

The self-consistency condition Eq.~(\ref{order_param}) can be used to
determine the critical temperature in the system. 
In order to determine the critical temperature $T_{c}$ one needs to set $\Delta\equiv 0$ in 
\begin{eqnarray}
\label{temp}
N(0)V\int_{\mu^*-\xi_{c}}^{\mu^*+\xi_{c}} d\xi\frac{\tanh (\frac{E_{n}+\eta }{2kT})+\tanh (%
\frac{E_{n}-\eta }{2kT})}{2 E}=1.
\end{eqnarray}%

The interesting property of the above equation is that for some values of
the asymmetry parameters it gives two positive critical temperatures. One
can interpret this result as the following. At zero and very low
temperatures the distributions of electrons and holes are step-like. The
paring in the case of a significant asymmetry will be suppressed, because
for different chemical potentials and carrier masses the particles will have
a considerable mismatch either in the energy or momentum. However, at some finite
temperature $T_{c1}$ the distributions of electrons and holes will become
smeared up enough due to thermal fluctuations, so there will be a
considerable probability of an electron and a hole to have matching energies
and momenta to create a Cooper pair. Eventually, at some higher temperature 
$T_{c2}$, thermal fluctuations become big enough to break Cooper pairs, and
for higher temperatures the pairing ceased to exist.
\begin{figure}[tbp]
\includegraphics[width=6.cm,angle=0]{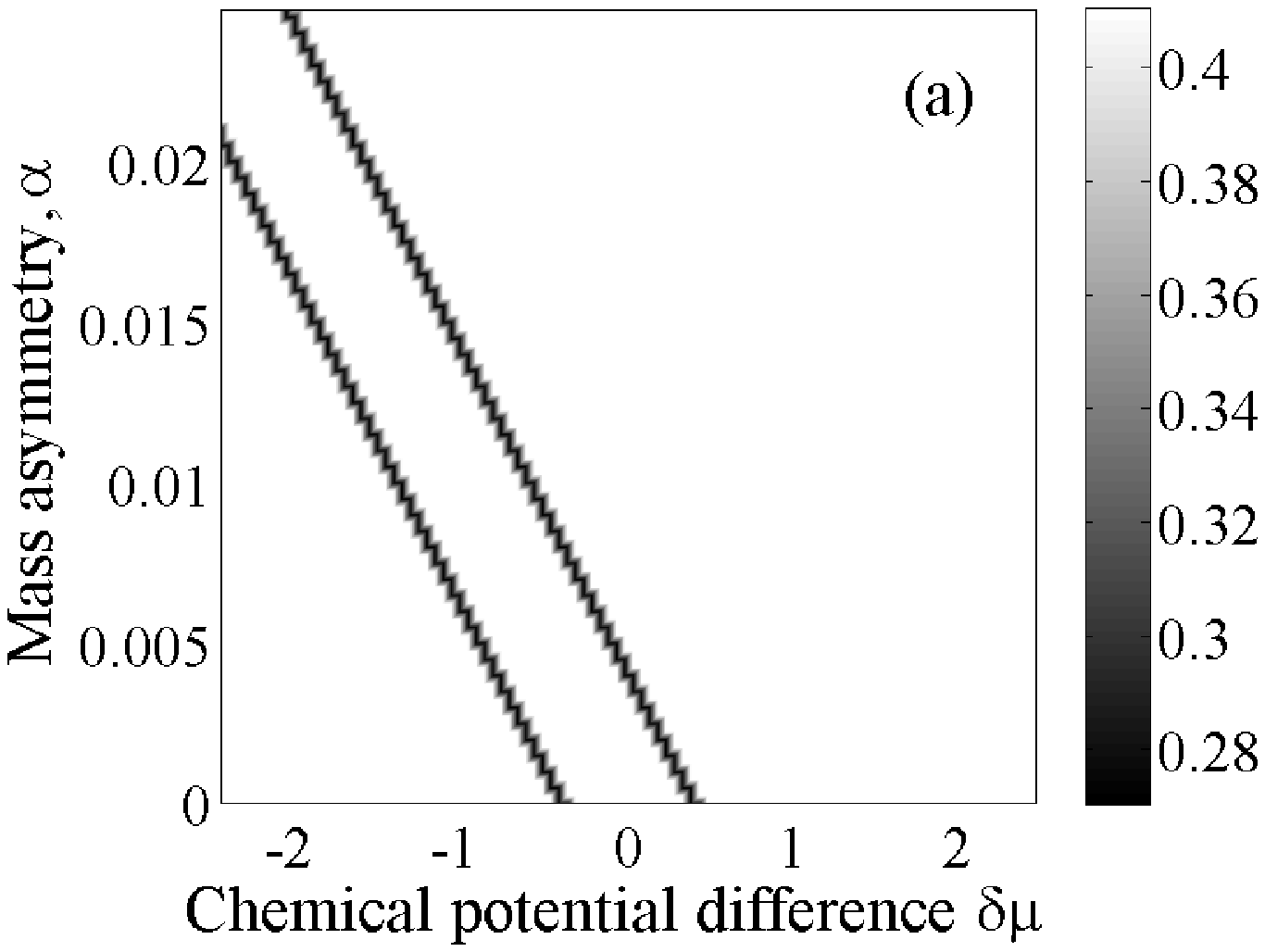} 
\includegraphics[width=6.cm,angle=0]{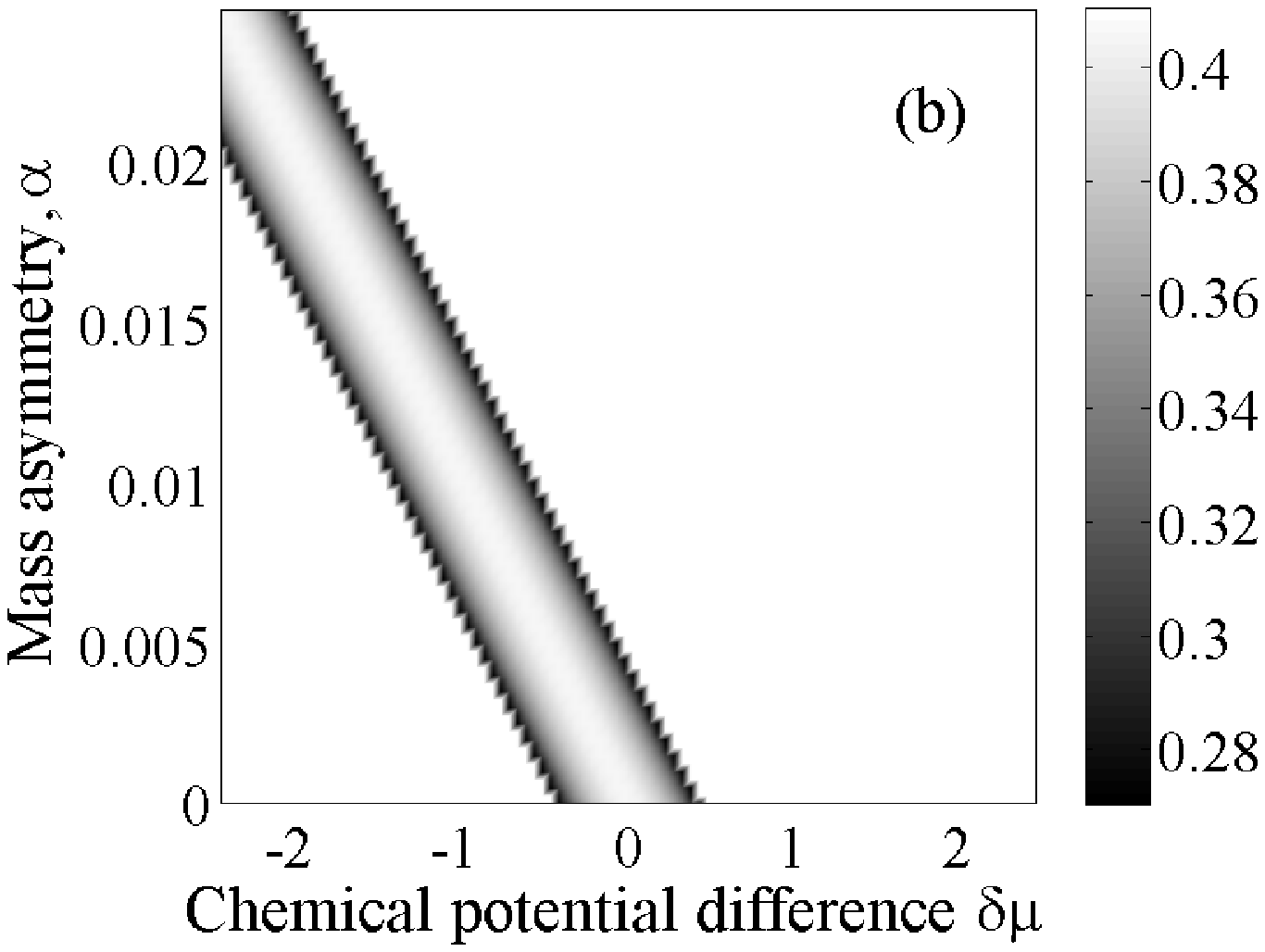}
\caption{{\protect\footnotesize {Critical temperature as a function of the
mass asymmetry parameter $\protect\alpha$ and the chemical potential
difference $\protect\delta \protect\mu$. For some parameters values Eq.~(\ref{temp}) 
for the critical temperature has two solutions: (a) the lower critical temperature $T_{c1}$;
(b) the upper critical temperature $T_{c2}$. Note that the condition $r\equiv
\protect\alpha\protect\mu^*+\delta \protect\mu=0$ corresponds to the maximum of the
upper critical temperature. System parameters are $\protect\xi_c/\protect\mu^*=0.1$%
, $V/\protect\mu^*=1.5\times 10^{-3}$. }}}
\label{label_figure1}
\end{figure}
In Fig.~\ref{label_figure1}(a) we plot the lower and  in Fig.~\ref{label_figure1}(b) the upper critical
temperatures as functions of the asymmetry parameters $\alpha $ and $\delta
\mu $.  The stripes in Fig.~\ref{label_figure1} correspond to the real-valued solutions of  Eq.~(\ref{temp}).   
The lower critical temperature $T_{c1}$ exists for a range of the asymmetry 
parameters $\alpha$ and $\delta \mu$, represented by two narrow bands shown in Fig.~\ref{label_figure1}(a), which are
both parallel to the line $r\equiv \alpha \mu ^{\ast }+\delta \mu =0$. The 
higher critical temperature $T_{c2}$ exists for the relatively broader range of the parameters, and is represented by 
one  wide band shown in Fig.~\ref{label_figure1}(b).  
Note that the condition $r=0$
corresponds to the maximum  of $T_{c2}$, when one asymmetry
effectively compensates the another one. This situation is very distinct
from the previous considerations of just one asymmetry parameter, where 
even tiny asymmetry in the system  adversely affects the superfluid state. 
In more general case of
two simultaneous  asymmetries, one can still achieve superfluidity, even for very
extreme values of the asymmetry parameters. The width of the
\textquotedblleft band\textquotedblright\ in the Fig.~1, where the superfluid
state is possible, is controlled by the magnitude of the gap $\Delta_{0}$
in the symmetric case.

In conclusion, we have studied the effect of electron-hole asymmetry on
formation of the superfluid state in a system of spatially separated electrons
and holes in two parallel planes. 
We predict that for some values of the electron-hole mass
asymmetry and mismatch of the chemical potentials two critical temperatures are
possible. The lower critical temperature corresponds to the superfluid
transition induced by thermal fluctuations making possible the pairing
between electrons and holes from distinct Fermi surfaces. 
We found that the mass
asymmetry can effectively compensate the mismatch between the chemical potentials
making the superfluid state possible in a much wider range of the system's
parameters than it was expected before. 
This effect can be used by
experimentalists to study superfluidity in a variety of asymmetric systems, including 
coupled quantum quantum wells.
It is found that for some asymmetry parameters the
normal and paired phases can coexist even at zero temperature.





\end{document}